\documentclass[conference]{IEEEtran}
\IEEEoverridecommandlockouts
\usepackage{cite}
\usepackage{amsmath,amssymb,amsfonts}
\usepackage{algorithmic}
\usepackage{graphicx}
\usepackage{textcomp}
\usepackage{xcolor}
\usepackage{subfigure}
\usepackage{caption}
\usepackage{booktabs}
\usepackage{multirow}
\usepackage{array}
\usepackage{float}
\def\BibTeX{{\rm B\kern-.05em{\sc i\kern-.025em b}\kern-.08em
    T\kern-.1667em\lower.7ex\hbox{E}\kern-.125emX}}
\newcommand\blfootnote[1]{%
  \begingroup
  \renewcommand\thefootnote{}\footnote{#1}%
  \addtocounter{footnote}{-1}%
  \endgroup
}

\newcommand{\beginsupplement}{%
        \setcounter{table}{0}
        \renewcommand{\thetable}{A\arabic{table}}%
        \setcounter{figure}{0}
        \renewcommand{\thefigure}{A\arabic{figure}}%
        \setcounter{section}{0}
        \renewcommand{\thesection}{A\arabic{section}}%
}

\makeatletter
\let\titleold\title
\renewcommand{\title}[1]{\titleold{#1}\newcommand{\thetitle}{#1}}
\def\maketitlesupplementary{%
  \newpage
  \twocolumn[%
    \begin{@twocolumnfalse}
      \centering
      {\Huge \thetitle\par}
      \vspace{0.2em}
      {\huge \textit{Supplementary Materials}\par}
      \vspace{1.0em}
      \begin{center}
      {\fontsize{11}{13.2}\selectfont
          Yi-Hsin Chen\textsuperscript{1} \quad
          Kuan-Wei Ho\textsuperscript{1} \quad
          Martin Benjak\textsuperscript{2} \quad
          Jörn Ostermann\textsuperscript{2} \quad
          Wen-Hsiao Peng\textsuperscript{1} \\[0.2em]
      }
      {\normalsize
        \textsuperscript{1}National Yang Ming Chiao Tung University, Taiwan \quad
        \textsuperscript{2}Leibniz Universität Hannover, Germany
      }
      \end{center}
      \vspace{1em}
    \end{@twocolumnfalse}
  ]%
}
\makeatother

\begin{document}

\title{Conditional Residual Coding with Explicit-Implicit Temporal Buffering for Learned Video Compression}

\author{%
\IEEEauthorblockN{
Yi-Hsin Chen\IEEEauthorrefmark{1} 
\quad Kuan-Wei Ho\IEEEauthorrefmark{1} 
\quad Martin Benjak\IEEEauthorrefmark{2}
\quad Jörn Ostermann\IEEEauthorrefmark{2}
\quad Wen-Hsiao Peng\IEEEauthorrefmark{1}
}
\IEEEauthorblockA{%
\IEEEauthorrefmark{1}National Yang Ming Chiao Tung University, Taiwan \quad 
\IEEEauthorrefmark{2}Leibniz Universität Hannover, Germany}
%\IEEEauthorblockA{%
%\tt\small \{yhchen12101.cs09@, kwho@cs., mick20001108.cs12@, abc900203abc.cs12@\}nycu.edu.tw \\ \tt\small wpeng@cs.nctu.edu.tw \tt\small alessandro.gnutti@unibs.it}
}

\maketitle

\begin{abstract}
This work proposes a hybrid,  explicit-implicit
temporal buffering scheme for conditional residual video coding. Recent conditional coding methods propagate implicit temporal information for inter-frame coding, demonstrating superior coding performance to those relying exclusively on previously decoded frames (i.e. the explicit temporal information). However, these methods require substantial memory to store a large number of implicit features. This work presents a hybrid buffering strategy. For inter-frame coding, it buffers one previously decoded frame as the explicit temporal reference and a small number of learned features as implicit temporal reference. Our hybrid buffering scheme for conditional residual coding outperforms the single use of explicit or implicit information. Moreover, it allows the total buffer size to be reduced to the equivalent of two video frames with a negligible performance drop on 2K video sequences. The ablation experiment further sheds light on how these two types of temporal references impact the coding performance. %\textcolor{red}{Experimental results demonstrate that using the buffer size of our hybrid buffering scheme for conditional residual coding can be reduced to the equivalent of two frames with negligible performance degradation in 2K video sequences.}

\end{abstract}


\vspace{-0.5cm}
\blfootnote{This work is supported by National Science and Technology Council, Taiwan, under the Grant NSTC 113-2634-F-A49-007- and National Center for High-performance Computing, Taiwan.}

\section{Introduction}
\label{sec:intro}
\begin{figure}[t]
\centering

\subfigure[Explicit temporal information buffering]{
\label{fig:fig1-a}
% \centering
\includegraphics[width=0.90\linewidth, trim= 0 190 0 5, clip]{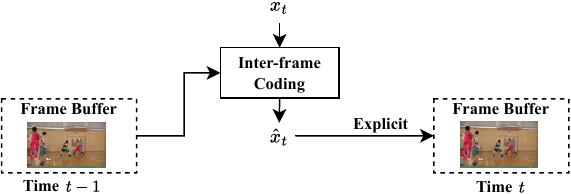}
}
\vspace{-3mm}
\subfigure[Implicit temporal information buffering]{
\label{fig:fig1-b}
% \centering
\includegraphics[width=0.90\linewidth, trim= 0 114 0 106, clip]{Figure/teaser/teaser_7.pdf} 
}
\vspace{-3mm}
\subfigure[Hybrid explicit and implicit temporal information buffering (ours)]{
\label{fig:fig1-c}
% \centering
\includegraphics[width=0.90\linewidth, trim= 0 7 0 184, clip]{Figure/teaser/teaser_7.pdf} 
} % 0 9 0 189
\caption{Comparison of different types of temporal information propagation for inter-frame coding.}
\vspace{-6mm}
\label{fig:teaser}
\end{figure}

% Explicit Temporal Buffering (Traditional codecs, & some learned video compression works)
Effectively leveraging information from previously decoded frames is pivotal for both traditional and learned video codec design. Similar to traditional codecs~\cite{avc,hevc,vvc}, many learned video codecs~\cite{dcvc, canf, maskcrt, mmsp24, mlvc,vlvc} explicitly buffer previously decoded frames in a decoded frame buffer, serving as the temporal reference information to assist with encoding the next frame (Fig.\ref{fig:teaser}~(a)). Essentially, these codecs can be viewed as recurrent neural networks with output-only recurrence, relying solely on decoded frames as the only contextual information from the past without maintaining or propagating any latent states temporally. Theoretically, this output recurrence design is less efficient, as decoded frames have to approximate the input coding frames while also having to provide a good summary of the past information.

% SOTA utilize Implicit Temporal Buffering (Another groups of learned video compression works, SOTA. Buffering size -> not practical)
In contrast to the methods that explicitly buffer decoded frames as the temporal reference information, another class of learned video codecs~\cite{tcm,hem,dcvc_dc,dcvc_fm, dcvcpqa, dcvcsdd, dcvclcg}, implicitly integrates and propagates the past temporal information by updating and buffering a large number of high-resolution features instead of the 3-channel decoded frames, as illustrated in Fig.\ref{fig:teaser}~(b). Since this large set of features is not constrained to approximate any input coding frames and consists of many channels, they are able to capture rich information from the past. %Although these methods still buffer one previously decoded frame, it is only used for motion estimation and does not directly provide contextual information for inter-frame coding. From the perspective of recurrent neural network design, this approach is closer to a design with recurrent hidden unit connections only. 
Although these methods achieve the state-of-the-art coding efficiency, they require substantial memory to store the large volume of features (e.g., equivalent to at least 21 frames in~\cite{tcm,hem} and 16 frames in\cite{dcvc_dc,dcvc_fm}). \textcolor{black}{As reported in~\cite{tcm}, reducing the number of buffered features to the equivalent of 3 frames results in a 2.5\% BD-rate increase. Therefore, subsequent works continue to buffer a large number of features.} \textcolor{black}{For reference, both HEVC~\cite{hevc} and VVC~\cite{vvc} only buffer 4 frames for predicted frame coding.}

%This results in high memory bandwidth demands for accessing the features from off-chip memory at high frame rates, posing a significant challenge for real-time applications on resource-constrained devices.

% Conditional Residual Coding vs. Conditional Coding 
In addition to making sure that the past information can be propagated efficiently, another critical aspect of designing a learned video codec is how the buffered temporal information is employed for inter-frame coding. The current mainstream approach is conditional coding~\cite{dcvc, canf, tcm, hem, dcvc_dc,dcvc_fm, dcvcpqa, dcvcsdd, dcvclcg} with which the buffered frames or features serve as condition signals for the inter-frame codec. It enables the non-linear utilization of the condition signals to encode the input frame. While state-of-the-art conditional coding shows promising results, a recent study~\cite{pcs22} discloses its potential information bottleneck issue. To alleviate this issue, Brand et al.~\cite{cond_res_coding} propose a conditional residual coding scheme that encodes the prediction residue $x_t - x_c$ using a conditional codec, where $x_t$ is the input frame and $x_c$ is the temporal predictor derived from the buffered temporal information. \cite{maskcrt, mmsp24} further demonstrate that conditional residual coding achieves superior coding performance to conditional coding. However, these experiments focus exclusively on scenarios that explicitly use a single reference frame for inter-frame coding, leaving largely unexplored the potential advantages of incorporating implicit temporal information.

In this work, we propose a hybrid temporal information buffering scheme for conditional residual coding. As illustrated in Fig.\ref{fig:teaser}~(c), our scheme explicitly buffers one previously decoded frame along with a small number of implicit features that represent additional temporal reference information from the past. Unlike the prior works~\cite{dcvc, canf, maskcrt,mmsp24, mlvc,vlvc} that rely solely on explicit temporal buffering, our hybrid buffering scheme is capable of leveraging more temporal reference information. In contrast to the approaches~\cite{tcm,hem,dcvc_dc,dcvc_fm, dcvcpqa, dcvcsdd, dcvclcg}, which primarily rely on implicit temporal information buffering, our method does not require buffering a large number of features. %From the perspective of recurrent neural network design, our hybrid scheme incorporates both output recurrence and hidden unit connections, which is theoretically more efficient than designs that solely use either output recurrence or hidden unit connections. 
The main contributions of this work are three-fold. (1) To the best of our knowledge, this is the first attempt in learned video compression to buffer temporal information both explicitly and implicitly within the framework of conditional residual coding. (2) Experimental results demonstrate that utilizing both explicit and implicit temporal information outperforms the sole use of any of them. (3) Compared to the state-of-the-art implicit buffering schemes for conditional coding, which require a large buffer size, the buffer size of our hybrid buffering scheme for conditional residual coding can be reduced to the equivalent of two frames \textcolor{black}{with negligible performance degradation in 2K video sequences.}

%In this work, we propose a hybrid temporal information buffering scheme for conditional residual coding. As illustrated in Fig.\ref{fig:teaser}~(c), our scheme explicitly buffers one previously decoded frame along with a small set of implicit features that carry additional temporal reference information. Unlike prior works~\cite{} that rely solely on explicit temporal information buffering, our hybrid buffering scheme is capable of leveraging more temporal reference information. In contrast to prior approaches~\cite{} that primarily rely on implicit temporal information buffering, our method does not require buffering a large volume of features. %From the perspective of recurrent neural network design, our hybrid scheme incorporates both output recurrence and hidden unit connections, which is theoretically more efficient than designs that solely use either output recurrence or hidden unit connections. 
%The main contributions of this work are two-fold. (1) To the best of our knowledge, this is the first attempt in learned video compression to buffer temporal information both explicitly and implicitly within the framework of conditional residual coding. (2) Experimental results demonstrate that the buffering size of our hybrid buffering scheme for conditional residual coding can be reduced to the equivalent of two frames with negligible performance degradation.

\section{Related Work}
\label{sec:related}
\subsection{Explicit and Implicit Temporal Information Buffering}
Based on the nature of the buffered temporal information for inter-frame coding, recent learned video compression works can be divided into two categories: explicit temporal information buffering and implicit temporal information buffering.

\textbf{Explicit temporal information buffering:} Methods in this category~\cite{dcvc, canf, maskcrt, mmsp24, mlvc,vlvc} explicitly buffer previously decoded frame(s) as reference information. \cite{dcvc, canf, maskcrt, mmsp24} use a single reference frame for inter-frame coding; however, the limited temporal information available from a single reference frame restricts their performance. To address this limitation, some works~\cite{mlvc,vlvc} \textcolor{black}{follow traditional codecs~\cite{avc,hevc,vvc} by buffering} multiple decoded frames and integrate them to construct a higher quality temporal predictor, which effectively improves the coding performance by utilizing more temporal information.

\textbf{Implicit temporal information buffering:} Unlike hand-crafted explicit buffering approaches, methods in this category~\cite{tcm, hem, dcvc_dc, dcvc_fm, dcvcpqa, dcvcsdd, dcvclcg} adopt a data-driven approach to learn and propagate temporal information in the feature domain. Some works~\cite{ dcvcpqa, dcvcsdd} employ convolutional long short-term memory (ConvLSTM) to preserve long-term temporal information, while others~\cite{tcm, hem, dcvc_dc, dcvc_fm} simplify the design by buffering intermediate features from the inter-frame decoder. To further exploit temporal information, \cite{dcvcpqa, dcvclcg} propose propagating two sets of features, one containing short-term information and the other containing long-term information.

\subsection{Conditional Coding and Conditional Residual Coding}
\textbf{Conditional coding:} Unlike traditional codecs~\cite{avc,hevc,vvc}, which adopt residual coding to encode pixel-domain residues between the input frame $x_t$ and its temporal predictor $x_c$, i.e., $x_t - x_c$, conditional coding~\cite{dcvc, canf, tcm, hem, dcvc_dc,dcvc_fm, dcvcpqa, dcvcsdd, dcvclcg} uses $x_c$ to condition the inter-frame codec in encoding the input frame $x_t$. From an information theory perspective, Ladune et al.\cite{mmsp} demonstrate that conditional coding is more efficient than traditional residual coding, as the conditional entropy $H(x_t|x_c)$ is smaller than or equal to the residual entropy $H(x_t - x_c)$. Building upon this, many recent works design their codecs around conditional coding and introduce new elements to improve coding performance, such as augmented normalizing flow-based framework~\cite{canf}, multi-scale temporal context conditioning~\cite{tcm}, advanced entropy models~\cite{hem, dcvc_dc}, improved temporal information modeling~\cite{dcvcpqa, dcvcsdd, dcvclcg}, offset diversity motion~\cite{dcvc_dc}, and refined training strategies~\cite{dcvc_dc, dcvc_fm}. However, despite showing promising results, conditional coding may suffer from the information bottleneck issue~\cite{pcs22} in practice, as some information from the temporal predictor $x_c$ can be lost during the feature extraction process, limiting the quality and effectiveness of the condition signal.

\textbf{Conditional residual coding:} To alleviate the bottleneck issue in conditional coding, Brand et al.~\cite{cond_res_coding} propose a conditional residual coding scheme that encodes the prediction resiude $x_t - x_c$ using a conditional codec. %They show that the conditional entropy rate $H(x_t - x_c | \tilde{x}_c)$ of the prediction residue is consistently lower than or equal to the conditional entropy rate $H(x_t | \tilde{x}_c)$, even in the presence of information bottlenecks, where $\tilde{x}_c$ denotes a processed version of $x_c$.
\textcolor{black}{They provide theoretical analyses showing that, in both lossless and lossy compression cases, conditional residual coding is at least as effective as conditional coding, even in the presence of information bottlenecks.} Building on this, Chen et al.\cite{maskcrt} further improve conditional residual coding by introducing a pixel-wise soft mask to switch between conditional coding and conditional residual coding. %Along this line of research, 
Chen et al.~\cite{mmsp24} further show that conditional residual coding offers higher coding performance with lower computational cost than conditional coding.

\section{Proposed Method}
\label{sec:method}
\subsection{System Overview}
\begin{figure*}[t!]
    \centering
    \includegraphics[width=0.845\linewidth, trim= 4 8 28 12, clip]{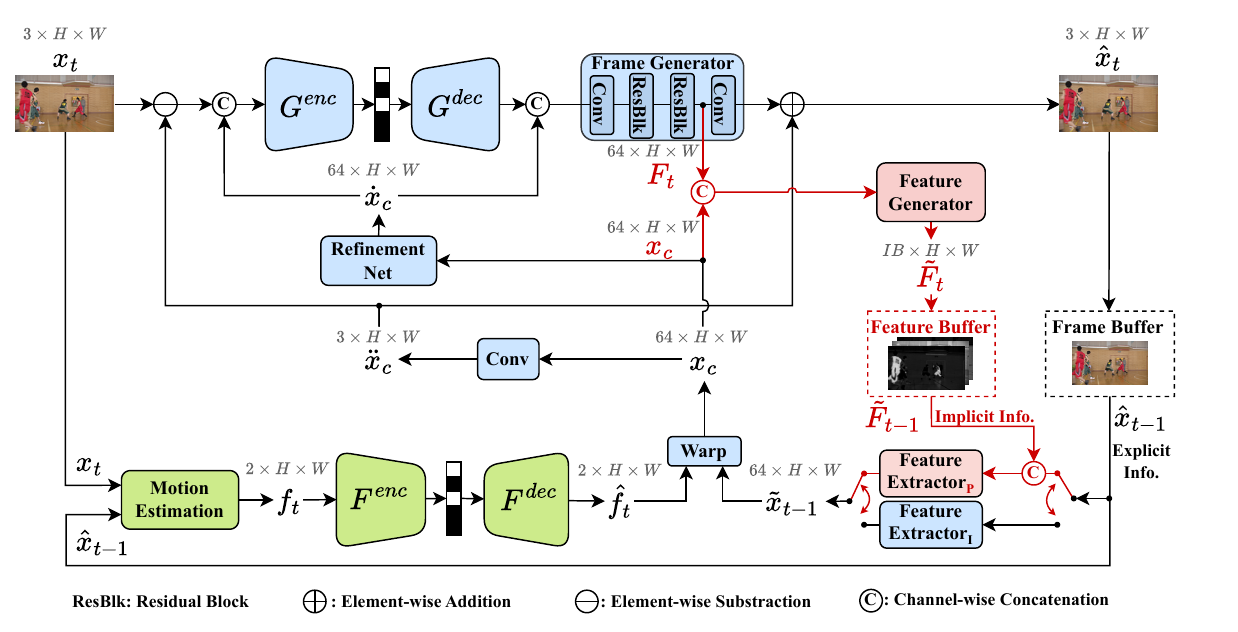}
    \vspace{-1mm}
    \caption{Overview of the proposed conditional residual coding framework with hybrid explicit and implicit temporal information buffering. The components highlighted in red represent the newly introduced elements compared to the conditional residual coding framework presented in~\cite{mmsp24}. }
    % \vspace{-5mm}
    \vspace{-4mm}
    \label{fig:overview}
\end{figure*}
Fig.~\ref{fig:overview} illustrates the main architecture of our proposed method. We use the conditional residual coding framework in~\cite{mmsp24} to explore the effectiveness of the hybrid buffering scheme and the impact of the buffer size on the coding performance. The red components in the figure highlight the newly introduced elements not in~\cite{mmsp24}.

The framework consists of a motion coding module (green-colored components) and an inter-frame coding module (blue- and red-colored components), along with two buffers (dashed-line boxes), namely a frame buffer and a feature buffer, which store explicit and implicit temporal reference information, respectively. The coding pipeline begins by estimating the motion between the input frame $x_t \in \mathbb{R}^{3 \times H \times W}$ and its reference frame $\hat{x}_{t-1} \in \mathbb{R}^{3 \times H \times W}$ to obtain an optical flow map $f_t \in \mathbb{R}^{2 \times H \times W}$, which is then encoded by the motion codec $\{F^{enc}, F^{dec}\}$. The decoded flow map $\hat{f}_t \in \mathbb{R}^{2 \times H \times W}$ is used to warp the temporal reference features $\tilde{x}_{t-1} \in \mathbb{R}^{64 \times H \times W}$, which are derived from the buffered explicit and implicit temporal reference information, to generate the temporal condition signal $x_c \in \mathbb{R}^{64 \times H \times W}$. As our inter-frame codec employs conditional residual coding, $x_c$ is used to obtain a pixel-domain temporal predictor $\ddot{x}_c \in \mathbb{R}^{3 \times H \times W}$ and a condition signal $\dot{x}_c \in \mathbb{R}^{64 \times H \times W}$ for the inter-frame codec. The inter-frame codec $\{G^{enc}, G^{dec}\}$ encodes the residue $x_t - \ddot{x}_c$ conditioned on $\dot{x}_c$, while on the decoder side, $\ddot{x}_c$ is added to the output of the frame generator to reconstruct the input frame. 

To assist with coding the next frame, we buffer not only the decoded frame $\hat{x}_t$, which serves as the explicit temporal reference information, but also $\tilde{F}_t \in \mathbb{R}^{IB \times H \times W}$, which provides the implicit temporal reference information by integrating $x_c$ and the intermediate feature $F_t \in \mathbb{R}^{64 \times H \times W}$ from the frame generator. Here, $IB$ represents the channel size of the buffered implicit temporal information. In this work, we adjust $IB$ to examine the impact of the buffer size on coding performance. Since none of $x_c$, $F_t$, $\tilde{F}_t$ is directly constrained to approximate the input frame $x_t$, and $\tilde{F}_t$ is updated and propagated over time, the implicit buffer is able to contain not only information from the current input frame but also from the previously coded frames, enabling the inter-frame codec to leverage more temporal reference information when coding the next frame.

\begin{figure*}[t]
    \centering
    \subfigure{
        \centering
        \includegraphics[width=.324\textwidth, trim= 0 0 55 50, clip]{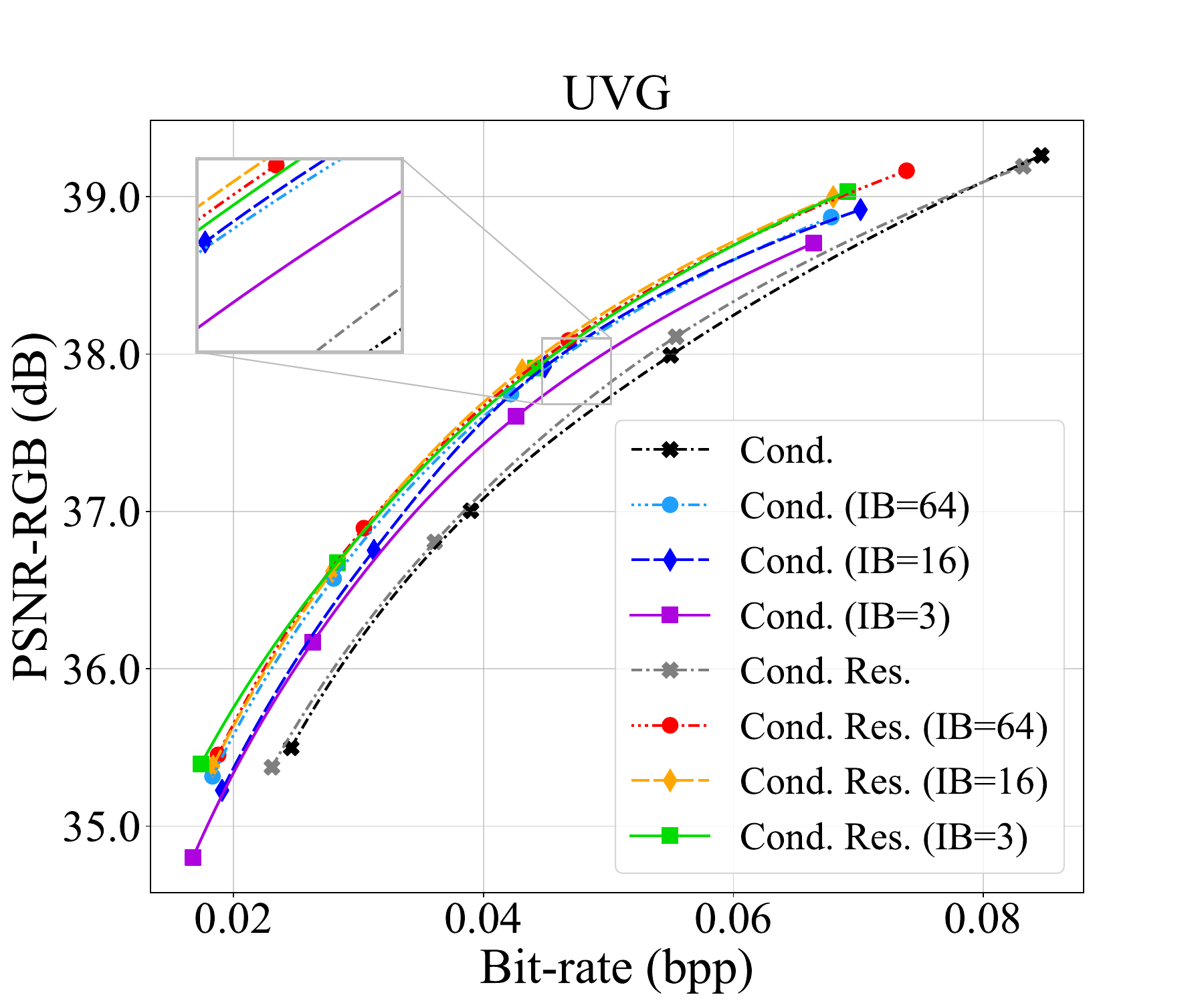}
        \label{fig:rd-a}
        }
    \hspace{-3.6mm}
    \subfigure{
        \centering
        \includegraphics[width=.324\textwidth, trim= 0 0 55 50, clip]{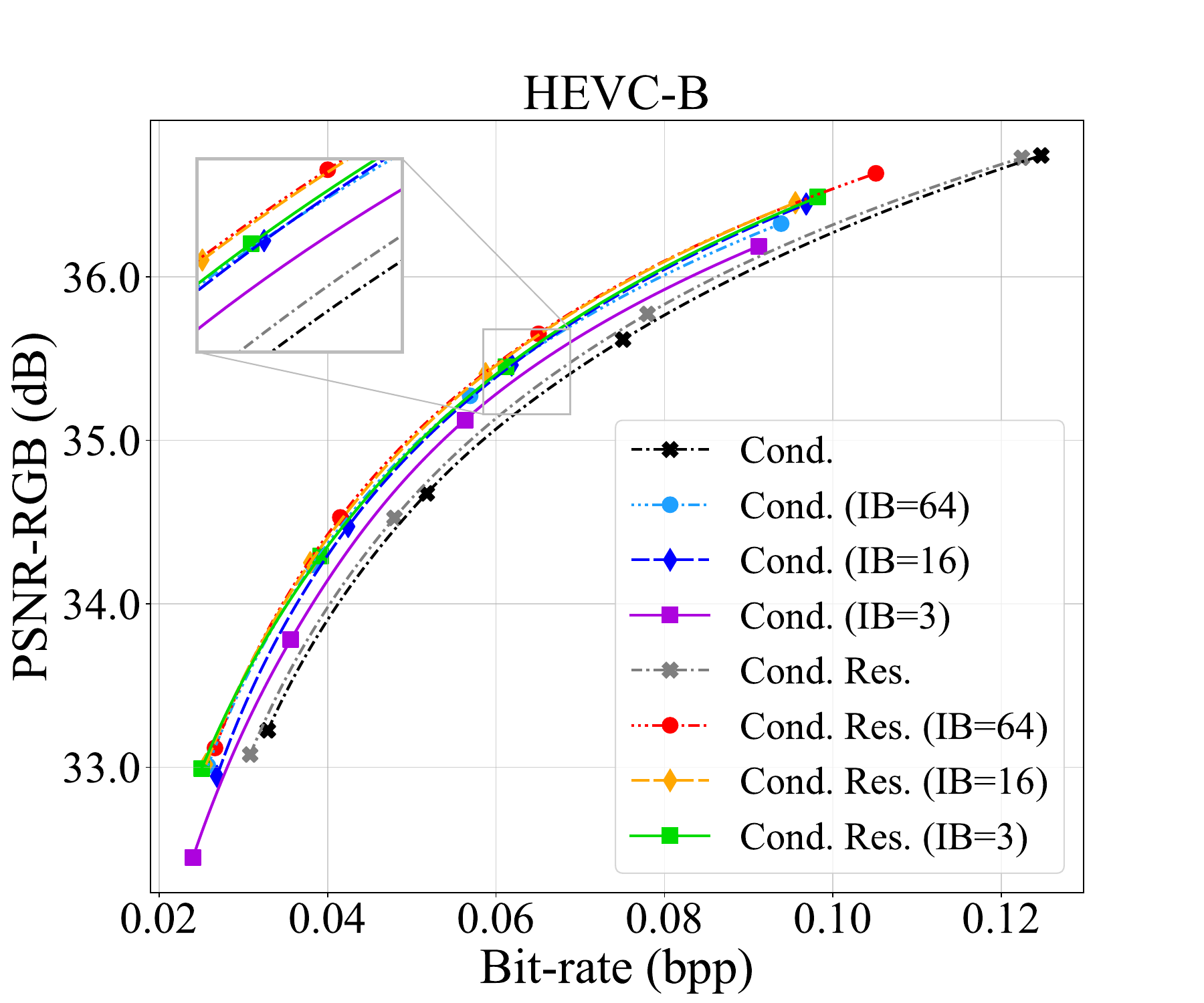}
        \label{fig:rd-b}
    }
    \hspace{-3.6mm}
    % \vspace{-1.1mm}
    \vspace{-0.2mm}
    \subfigure{
        \centering
        \includegraphics[width=.324\textwidth, trim= 0 0 55 50, clip]{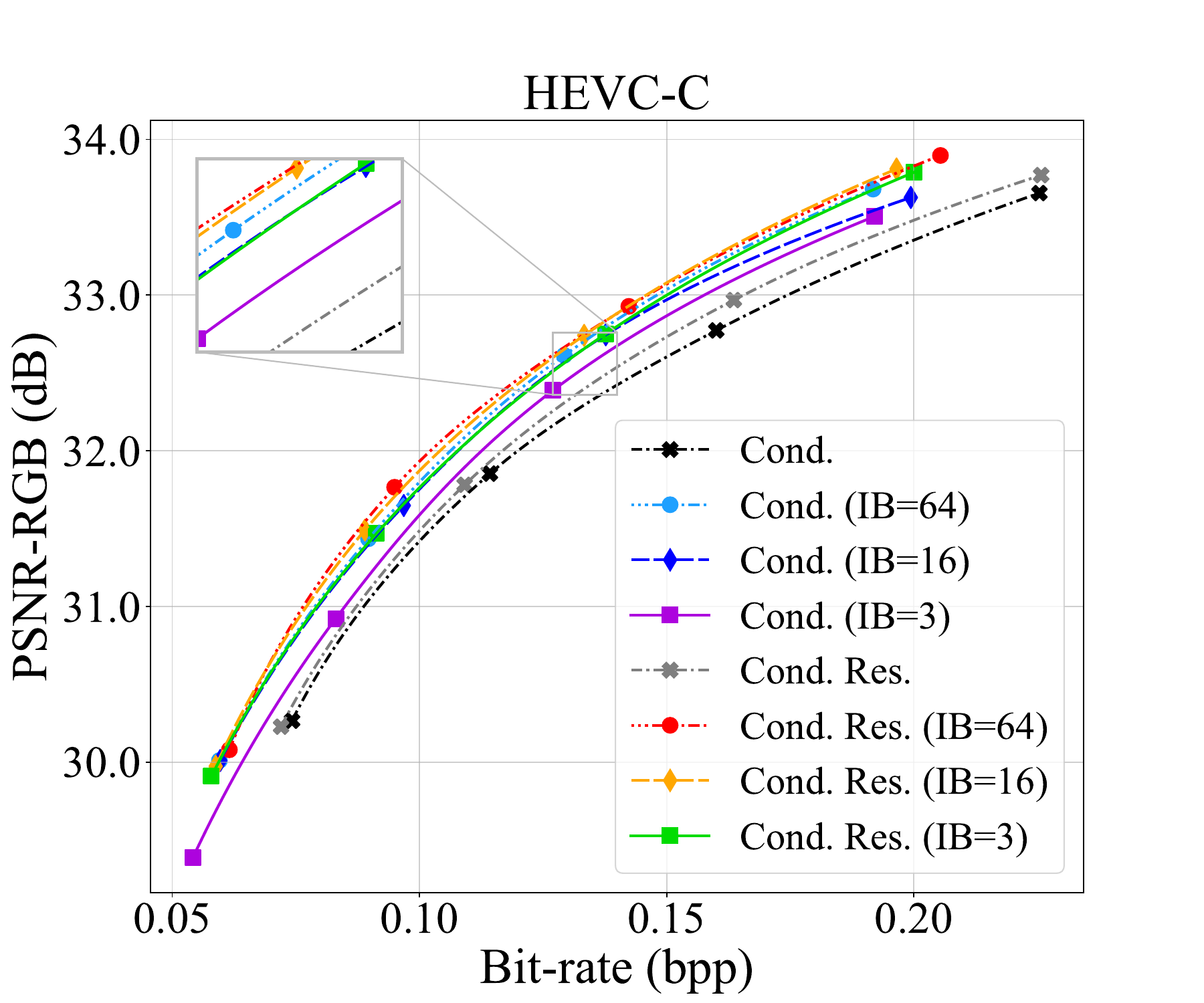}
        \label{fig:rd-c}
    }
    \subfigure{
        \centering
        \includegraphics[width=.324\textwidth, trim= 0 0 55 50, clip]{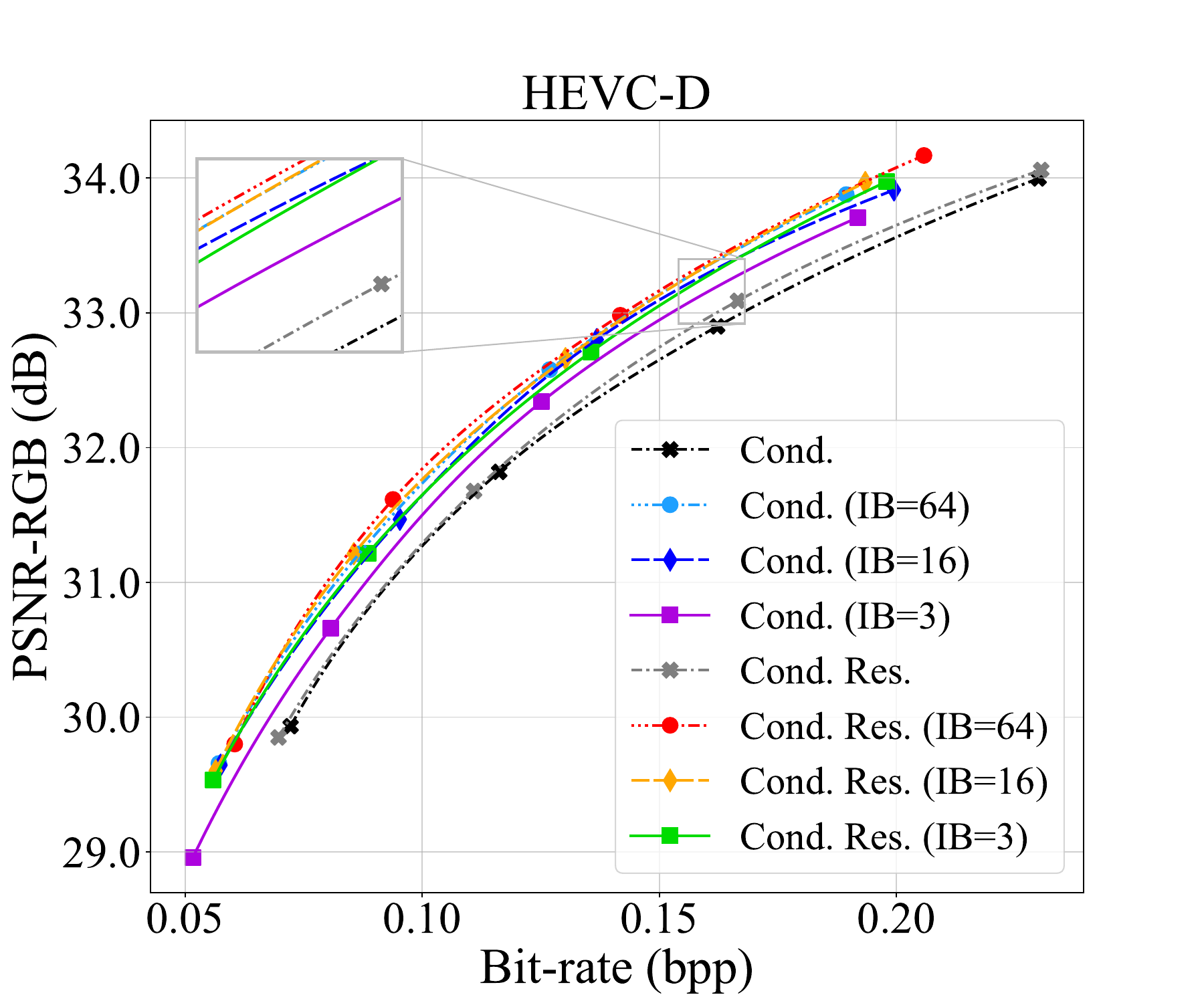}
        \label{fig:rd-d}
        }
    \hspace{-3.6mm}
    \subfigure{
        \centering
        \includegraphics[width=.324\textwidth, trim= 0 0 55 50, clip]{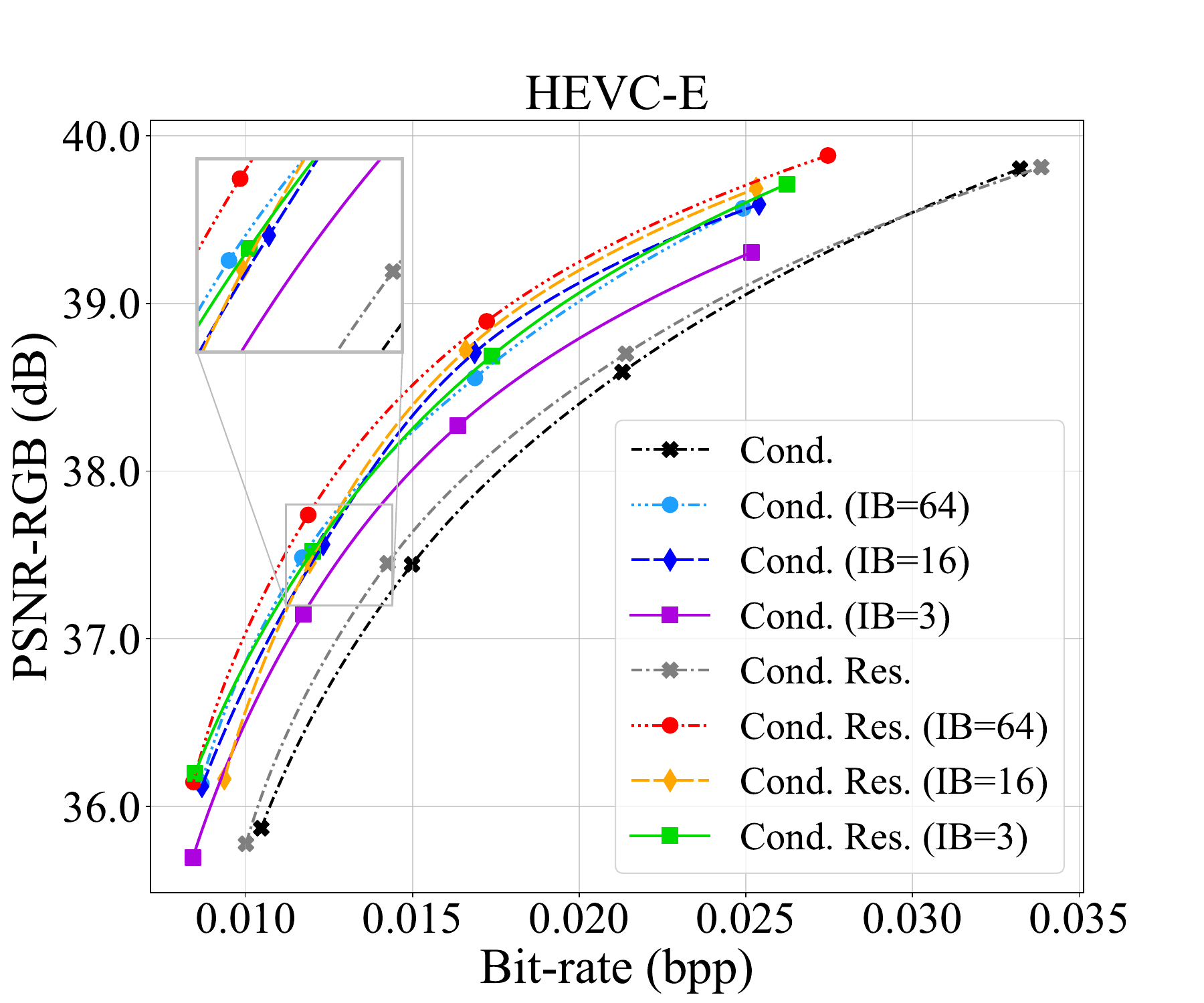}
        \label{fig:rd-e}
    }
    \hspace{-3.6mm}
    \subfigure{
        \centering
        \includegraphics[width=.324\textwidth, trim= 0 0 55 50, clip]{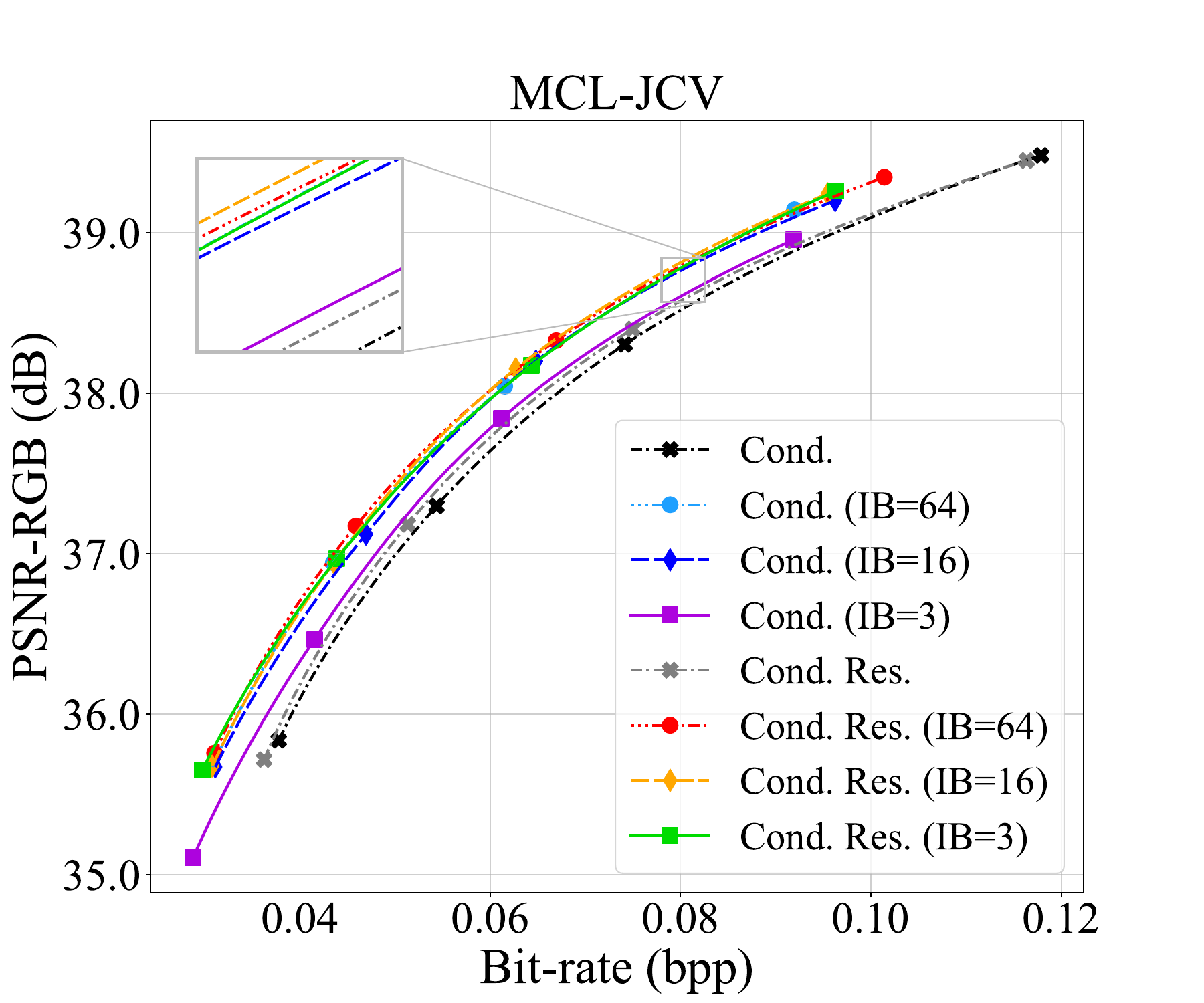}
        \label{fig:rd-f}
    }
    % \vspace{-6.1mm}
    \vspace{-4.5mm}
    \caption{Rate-distortion comparison between conditional coding and conditional residual coding with varying buffer sizes.}
    \vspace{-0.1cm}
    \label{fig:main_RD}
\end{figure*}
\begin{table*}[]
\caption{BD-rate (\%) comparison in terms of PSNR-RGB. The anchor is conditional coding without using implicit temporal information. The values in parentheses indicate the BD-rate changes relative to the codec type with $IB=64$.}
\label{table:main_RD}
\centering
\begin{tabular}{l|cccccc|c}
\toprule
                        & UVG & HEVC-B & HEVC-C & HEVC-D & HEVC-E & MCL-JCV & Average \\ 
% Only F_t
\midrule
    Cond.               & 0         & 0         & 0      & 0       & 0       & 0       & 0 \\
    Cond. (IB=64)       & -16.36    & -13.38    & -12.34  & -12.97  & -21.11  & -14.43  & -15.10  \\
    % Cond. (IB=16)       & -14.97 (+1.39) & -12.39 (+0.99) & -11.83 (+0.51) & -12.56 (+0.41) & -21.66 (-0.55) & -14.17 (+0.26)  & -14.60 (+0.50) \\
    Cond. (IB=16)       & -14.58 (+1.78) & -11.83 (+1.55) & -11.04 (+1.30) & -11.33 (+1.64) & -21.23 (-0.12) & -13.42 (+1.01)  & -13.91 (+1.19) \\
    Cond. (IB=3)        & -9.26 (+7.10) & -7.96 (+5.42) & -6.44 (+5.90) & -6.93 (+6.04) & -14.56 (+6.55)  & -7.11 (+7.32)  & -8.71 (+6.39) \\
    % Cond. (IB=3)        & -14.29 (+2.07) & -10.3 (+3.08) & -8.84 (+3.50) & -9.3 (+3.67) & -17.35 (+3.76)  & -10.66 (+3.77)  & -11.79 (+3.31) \\
\midrule
    Cond. Res.          & -1.66     & -2.57     & -3.07  & -1.55  & -3.55    & -3.53  & -2.66  \\
    Cond. Res. (IB=64)  & -19.06    & -15.18    & -14.65 & -14.25  & -26.05  & -15.89  & -17.51  \\
    Cond. Res. (IB=16)  & -18.95 (+0.11) & -15.01 (+0.17) & -14.10 (+0.55) & -13.36 (+0.89) & -21.35 (+4.70) & -14.10 (+1.79) & -16.15 (+1.37) \\
    Cond. Res. (IB=3)   & -18.17 (+0.89) & -13.77 (+1.41) & -11.86 (+2.79) & -10.97 (+3.28) & -21.37 (+4.68) & -14.02 (+1.87) & -15.03 (+2.49) \\

\bottomrule
\end{tabular}
\vspace{-3mm}
\end{table*}

\subsection{Implicit Temporal Information Buffering}
Following the implicit temporal buffering works~\cite{tcm,hem,dcvc_dc,dcvc_fm, dcvcpqa, dcvcsdd, dcvclcg}, we leverage the information-rich features, $F_t \in \mathbb{R}^{64 \times H \times W}$, from the frame generator before reconstruction to construct the implicit temporal information for the subsequent coding frame. To investigate the impact of the buffer size on coding performance, we introduce a feature generator to adjust the channel size $IB$ of $F_t$. 

The architecture of the feature generator is identical to that of the frame generator, except for the channel sizes of the first and last convolutional layers, which are modified to match the input and output channel sizes, respectively. Furthermore, unlike the previous methods~\cite{tcm,hem,dcvc_dc,dcvc_fm, dcvcpqa, dcvcsdd, dcvclcg} that directly buffer these features as the implicit temporal reference information, we also introduce the warped features $x_c$ as the input to the feature generator. The resulting fused features, with a reduced channel size, are then buffered and used as the implicit temporal reference information for the next coding frame. This design feature stems from the fact that the prior works~\cite{tcm,hem,dcvc_dc,dcvc_fm, dcvcpqa, dcvcsdd, dcvclcg} are based on conditional coding frameworks, where their $F_t$ contains substantial contextual information from the input frame $x_t$. In contrast, this work adopts a conditional residual coding framework, where the input of the inter-frame codec is the residue $x_t - \ddot{x}_c$, making $F_t$ naturally less information-rich. Consequently, we incorporate the temporal predictor $x_c$ in buffering the contextual information. 

\subsection{Explicit and Implicit Temporal Information Fusion}
To minimize the changes to the base codec and ensure a fair comparison, we follow the design in \cite{mmsp24}, which uses a single set of reference features for warping to obtain the temporal predictor $x_c$. Unlike the codec in~\cite{mmsp24}, which adopts $\hat{x}_{t-1}$ as the only reference, our approach employs two references: the explicit temporal reference $\hat{x}_{t-1}$ and the implicit temporal reference $\tilde{F}_{t-1}$. These references are concatenated along the channel dimension and fused by a feature extractor, denoted as Feature Extractor$_P$ in Fig.~\ref{fig:overview}, to obtain the temporal predictor $x_c$. The architecture of Feature Extractor$_P$ is identical to the feature extractor in~\cite{mmsp24}, except for an adjustment to the channel size of the first convolutional layer in order to accommodate the concatenated input. Notably, for the first predicted frame (P-frame), where the previous frame is intra coded and the implicit temporal reference is unavailable, we use a separate feature extractor, denoted as Feature Extractor$_I$ in Fig.~\ref{fig:overview}, which adopts the previously decoded frame as the only input.

\section{Experiments}
\label{sec:experiment}
\subsection{Settings}
\textbf{Training details:} We train our models on the Vimeo-90k dataset~\cite{vimeo} with the sequences randomly cropped into $256 \times 256$ patches. The model is trained by initializing with pre-trained base codec weights. The feature generator is first optimized with a 3x3 convolution to map its output to the RGB domain, regularized with the coding frame. The remaining training procedure is the same to \cite{mmsp24}. We learn four separate models, with $\lambda$ set to $\{256, 512, 1024, 2048\}$ in the training objective, where $\lambda$ is a hyper-parameter that controls the trade-off between distortion and rate. The distortion is quantified by the mean squared reconstruction error in the RGB domain.

\textbf{Baseline methods:} We compare our method with the base codec, which explicitly buffers only the previous decoded frame. We also compare our method with a conditional coding framework that buffers both the explicit and implicit temporal information with a variable buffer size. For a fair comparison, we adapt the conditional coding framework in \cite{mmsp24} by introducing a feature generator to adjust the size of the buffer for storing the implicit temporal information and a Feature Extractor$_P$ to fuse the buffered explicit and implicit temporal information. This baseline method with conditional coding has nearly the same coding components as ours, except that our inter-frame codec employs conditional residual coding. With this baseline method, the input to the feature generator is $F_t$, which follows the idea of the state-of-the-art implicit buffering approaches~\cite{tcm,hem,dcvc_dc,dcvc_fm, dcvcpqa, dcvcsdd, dcvclcg} for conditional coding.

\begin{figure}[t!]
    \centering
    \includegraphics[width=0.96\linewidth]{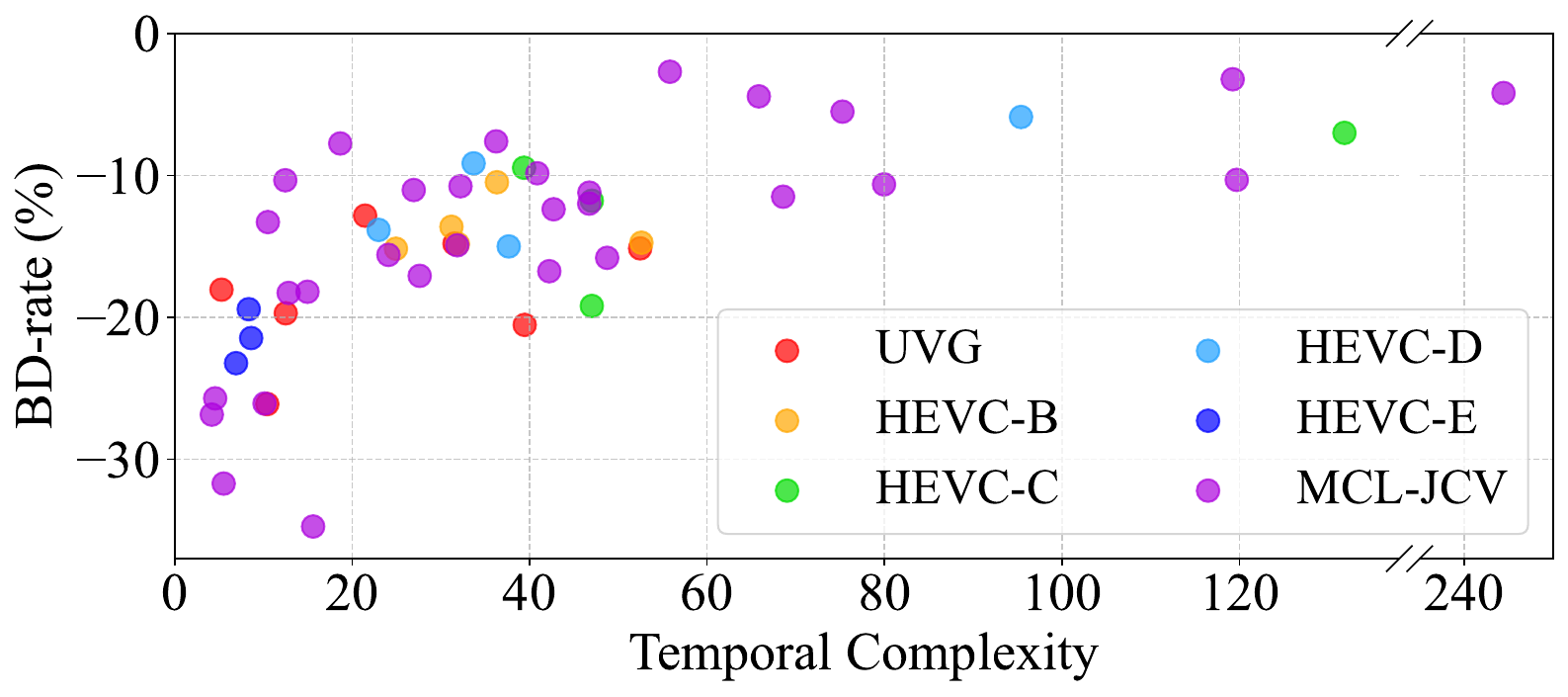}
    % \vspace{-2mm}
    \vspace{-1mm}
    \caption{Analysis of BD-rate versus temporal complexity for conditional residual coding with $IB=3$, using conditional residual coding without implicit temporal information as the anchor. Each point represents the result of a single test sequence. Temporal complexity is calculated using the video complexity analyzer from \cite{vca}.}
    % \vspace{-2mm}
    \vspace{-1mm}
    \label{fig:per_seq}
\end{figure}

\begin{table}[t]
\caption{Comparison of the BD-rate and complexity in terms of the encoding/decoding MACs, model size and the required buffer size of the full-resolution feature maps.}% for inter-frame coding.}
\label{table:complexity}
\centering
\setlength{\tabcolsep}{4pt}
\begin{tabular}{l|ccccc}
\toprule
    & \begin{tabular}[c]{@{}c@{}} BD-rate \\ (\%)  \end{tabular} 
    & \begin{tabular}[c]{@{}c@{}} Encoding / Decoding  \\ kMACs/pixel \end{tabular} 
    & \begin{tabular}[c]{@{}c@{}} Model \\ Size (M) \end{tabular} 
    & \begin{tabular}[c]{@{}c@{}} Buffer \\ Size \end{tabular} \\ 
\midrule
    Cond.               &   0    & 1153 / 762 & 7.944 &  3 \\
    Cond. (IB=64)       & -15.10 & 1375 / 984 & 8.279 & 67 \\
    % Cond. (IB=16)       & -14.60 & 1348 / 957 & 8.223 & 19 \\
    Cond. (IB=16)       & -13.91 & 1348 / 957 & 8.223 & 19 \\
    Cond. (IB=3)        & -8.71  & 1340 / 949 & 8.208 &  6 \\
\midrule
    Cond. Res.          & -2.66  & 1155 / 764  & 7.946 &  3 \\ 
    Cond. Res. (IB=64)  & -17.51 & 1451 / 1060 & 8.317 & 67 \\
    Cond. Res. (IB=16)  & -16.15 & 1395 / 1004 & 8.262 & 19 \\ 
    Cond. Res. (IB=3)   & -15.03 & 1380 / 989  & 8.247 &  6 \\
\bottomrule
\end{tabular}
\vspace{-4mm}
\end{table}

% \textbf{Evaluation:} We evaluate our method on several widely used test datasets, including UVG~\cite{uvg}, %HEVC Class B~\cite{hevcctc}, HEVC Class C~\cite{hevcctc}, HEVC Class D~\cite{hevcctc}, HEVC Class E~\cite{hevcctc}, 
% HEVC Class B $\sim$ E~\cite{hevcctc}, and MCL-JCV~\cite{mcl}. 
Several widely used test datasets, including UVG~\cite{uvg}, %HEVC Class B~\cite{hevcctc}, HEVC Class C~\cite{hevcctc}, HEVC Class D~\cite{hevcctc}, HEVC Class E~\cite{hevcctc}, 
HEVC Class B $\sim$ E~\cite{hevcctc}, and MCL-JCV~\cite{mcl}, are used for evaluation. Following \cite{dcvc_dc, mmsp24}, all the YUV420 test sequences are first converted to RGB444 using BT.709~\cite{ffmpeg}, followed by encoding the first 96 frames in each test sequence. The intra-period is 32. The BD-rate savings are reported in terms of Peak Signal-to-Noise Ratio (PSNR) in the RGB domain. The bit rate is in bits per pixel (bpp). Negative and positive BD-rate numbers suggest rate reduction and inflation, respectively. Following the common test protocol of traditional codecs~\cite{avc,hevc,vvc}, the average BD-rate of a dataset is obtained by averaging the BD-rate savings over individual sequences in the dataset.

% , HEVC-RGB~\cite{hevcrgb}

%calculated using piecewise cubic interpolation~\cite{interpolation}. 

\subsection{Experimental Results}
Fig.~\ref{fig:main_RD} presents the rate-distortion performance comparison and Table~\ref{table:main_RD} reports its corresponding BD-rate numbers. We adjust the channel size $IB$ of the buffered implicit features to assess the impact of buffer size on coding performance. The following observations can be made.

(1) The additional use of the implicit temporal information significantly improves coding performance in both conditional and conditional residual coding. The performance improvement is especially notable on the HEVC-E dataset, likely due to its video conferencing content with static backgrounds. Buffering implicit temporal information enables the use of higher-quality references, such as intra frame information, which contributes to better coding performance. \textcolor{black}{Fig.~\ref{fig:per_seq} further presents how the BD-rate savings of individual test sequences are correlated with their temporal complexity. As shown, leveraging the implicit temporal information yields higher gains in test sequences with lower temporal complexity. This result is in line with the higher coding performance on the HEVC-E dataset.}

(2) Conditional residual coding consistently outperforms conditional coding across all the buffer sizes. Interestingly, conditional residual coding with $IB=3$ achieves comparable or even better coding performance than conditional coding with $IB=64$.

(3) On 2K test sequences (i.e., UVG, HEVC-B, and MCL-JCV datasets), the feature buffer size of our hybrid buffering scheme for conditional residual coding can be reduced from 64 channels to 3 channels with modest performance degradation (a BD-rate increase of less than 2\%). %\textcolor{black}{Note that buffering a 3-channel implicit feature and a decoded frame requires a buffer size equivalent to buffering two frames.} 
In contrast, conditional coding is more sensitive to buffer size on 2K sequences; reducing the feature buffer size from 64 channels to 3 channels leads to 5\%-7\% performance drops.

Table~\ref{table:complexity} analyzes how the BD-rate saving varies with the model's complexity characterized by the kMAC/pixel, model size, and buffer size. We see that introducing the implicit temporal information (i.e. the variants with IB 64, 16, or 3) significantly improves coding efficiency but also increases complexity due to the need to process this additional information. We note that both the kMAC/pixel and model size can be further reduced through network optimization. However, the buffer size is a design choice. 

%this is not the main focus of this work and is among our future work. %Comparing conditional coding with $IB=64$ and conditional residual coding with $IB=3$, both have similar kMAC/pixel and model sizes and performance, but the latter requires a smaller buffer size. Note that larger buffer sizes result in higher memory bandwidth demands when accessing features from off-chip memory at high frame rates, which presents a significant challenge for real-time applications on resource-constrained devices.

\begin{table}[t]
\caption{Explicit vs. implicit vs. hybrid temporal information buffering. The anchor is our conditional residual codec with a 3-channel $\hat{x}_{t-1}$ as the only explicit information.}
\vspace{-1mm}
\label{table:hybrid}
\centering
\setlength{\tabcolsep}{10pt}
\begin{tabular}{l|cc|cc}
\toprule
    & \begin{tabular}[c]{@{}c@{}} Implicit \\ (IB=67) \end{tabular} 
    & \begin{tabular}[c]{@{}c@{}} Hybrid \\ (IB=64) \end{tabular}
    & \begin{tabular}[c]{@{}c@{}} Implicit \\ (IB=6) \end{tabular} 
    & \begin{tabular}[c]{@{}c@{}} Hybrid \\ (IB=3) \end{tabular} \\
\midrule 
    UVG  & -5.72 & -17.76 & 0.45 & -16.89   \\
    HEVC-B  & -4.91 & -13.07 & 0.62 & -11.67   \\
    HEVC-C  & -3.65 & -12.09 & -1.12 & -9.33  \\
    HEVC-D  & -4.60 & -12.96 & -2.41 & -9.71   \\
    HEVC-E  & -4.28 & -23.29 & 10.33 & -18.32   \\
    MCL-JCV  & -3.07 & -12.80 & 2.90 & -10.74   \\
\midrule 
    Average   & -4.37 & -15.33 & 1.80 & -12.78   \\
\bottomrule
\end{tabular}
\vspace{-3mm}
\end{table}

\subsection{Ablation Study}
\textbf{Hybrid temporal information buffering:} Table~\ref{table:hybrid} presents an ablation study comparing the hybrid of both explicit and implicit temporal reference information with their single use, i.e., either $\hat{x}_{t-1}$ or $\tilde{F}_{t-1}$ in Fig.~\ref{fig:overview}. Note that the buffer size for using only implicit temporal information is set to 67 or 6 to ensure a fair comparison with our hybrid buffering scheme, which buffers a 64-channel or a 3-channel $\tilde{F}_{t-1}$ along with a 3-channel $\hat{x}_{t-1}$. %which buffers a 64-channel $\tilde{F}_{t-1}$ along with a 3-channel $\hat{x}_{t-1}$, or a 3-channel $\tilde{F}_{t-1}$ along with a 3-channel $\hat{x}_{t-1}$. 
We choose the anchor to be our conditional residual coding framework with a 3-channel $\hat{x}_{t-1}$ as the only explicit information. 

%From Table~\ref{table:hybrid}, we observe 
Table~\ref{table:hybrid} shows that both implicit and hybrid schemes with large buffer sizes outperform the explicit scheme (with only a 3-channel $\hat{x}_{t-1}$). It is expected that having more channels allows more temporal reference information to be stored for contextual coding. %It is expected that having more channels allows more temporal information from the past to be stored for contextual coding. 
Interestingly, when the implicit buffer size is reduced to 3, our hybrid scheme outperforms the implicit variant with IB=6. It suggests that the explicit $\hat{x}_t$ serves as a strong reference, which need not be learned. In comparison, using only implicit information with a 6-channel buffer performs slightly worse than the explicit scheme with only a 3-channel $\hat{x}_{t-1}$. We conjecture that it takes more effort to learn well the implicit information that is critical to contextual coding. The training strategy must be delicately crafted.

%This may be due to the challenge of learning in a data-driven manner to balance key information from the previous frame with temporal information from earlier frames under a limited buffer size.

\begin{table}[t]
\caption{Ablation on the input to the feature generator. The anchor is our conditional residual coding framework with a 3-channel $\hat{x}_{t-1}$ as the only explicit information.}
\label{table:generator}
\centering
\setlength{\tabcolsep}{5.5pt}
\begin{tabular}{l|cccccc}
\toprule
    & \multicolumn{3}{c}{Hybrid (IB=64)} & \multicolumn{3}{c}{Hybrid (IB=3)} \\
    \cmidrule(lr){2-4} \cmidrule(lr){5-7}
    &  $x_c$    & $F_t$    & $x_c \& F_t$   & $x_c$    & $F_t$    & $x_c \& F_t$  \\
\midrule
    UVG         & -12.57 & -13.22 & -17.76 & -13.53 & -9.98 & -16.89  \\
    HEVC-B      & -7.54 & -10.40 & -13.07 & -6.09 & -8.65 & -11.67  \\
    HEVC-C      & -7.87 & -8.77 & -12.09 & -5.61 & -8.29 & -9.33  \\
    HEVC-D      & -8.86 & -10.00 & -12.96 & -6.80 & -8.28 & -9.71  \\
    HEVC-E      & -14.07 & -16.00 & -23.29 & -10.34 & -13.28 & -18.32  \\
    MCL-JCV     & -6.02 & -8.68 & -12.80 & -4.39 & -6.27 & -10.74  \\
\midrule
    Average     & -9.49 & -11.18 & -15.33 & -7.79 & -9.13 & -12.78  \\
\bottomrule
\end{tabular}
\vspace{-3mm}
\end{table}

\textbf{Input to the feature generator:} Table~\ref{table:generator} presents an ablation study comparing the use of both $F_t$ and $x_c$ as inputs to the feature generator with the single use of only one of them. As shown, using either $F_t$ or $x_c$ as the feature generator input results in a significant performance gain compared to the anchor that does not utilize implicit temporal information, and using both $F_t$ and $x_c$ yields the highest coding gain. These results further confirm the effectiveness of propagating both explicit and implicit temporal information. Notably, using $F_t$ only performs better than the variant with $x_c$ only. %\textcolor{black}}{This may be attributed to the fact that $F_t$ has more information about the current coding frame than $x_c$. That extra information is decoded from the compressed bitstream (see Fig.~\ref{fig:overview}).}
\textcolor{black}{This is because the process of generating $F_t$ allows access to more information about the current coding frame signaled in the bitstream, enabling the model to effectively compare the current information with the past information (provided by $\dot{x}_c$) and select the most critical temporal information for buffering. In contrast, $x_c$ can only access limited information from the decoded motion.}

% Notably, using $F_t$ alone performs better than using $x_c$ alone because $F_t$ can access information from the current coding frame transmitted through the residual bitstream, enabling a more effective comparison with the past information provided by $\dot{x}_c$ to select the most critical temporal information for buffering, while $x_c$ can only obtain limited information from the decoded motion.
\section{Conclusion}
\label{sec:conclusion}
In this work, we buffer one previously decoded frame as explicit temporal reference along with few implicit features that provide additional temporal reference for conditional residual coding. Experimental results confirm the superiority of using hybrid temporal information over relying on either explicit or implicit temporal information alone. Furthermore, the buffer size of our hybrid scheme for conditional residual coding can be reduced to the equivalent of two video frames only with minimal performance degradation on 2K video sequences. Extending our method to state-of-the-art learned video compression models is among our future work.

\bibliographystyle{IEEEbib}
\bibliography{icme2025references}

% % \documentclass[conference]{IEEEtran}
% \documentclass[onecolumn]{IEEEtran}
% \IEEEoverridecommandlockouts
% % The preceding line is only needed to identify funding in the first footnote. If that is unneeded, please comment it out.
% \usepackage{cite}
% \usepackage{amsmath,amssymb,amsfonts}
% \usepackage{algorithmic}
% \usepackage{graphicx}
% \usepackage{textcomp}
% \usepackage{xcolor}
% \usepackage{subfigure}
% \usepackage{booktabs}
% \usepackage{multirow}
% \usepackage{array}
% \usepackage{caption}
% % \usepackage{subcaption}
% \usepackage{float}
% \usepackage{placeins}
% \usepackage{multicol}

% % \usepackage{graphicx}
% % \usepackage{subcaption} % Required for subfigure support

% % Modify subfigure captions to show only (a), (b), (c)
% \captionsetup[subfigure]{labelformat=simple, labelsep=none}
% \renewcommand\thesubfigure{(\alph{subfigure})} % To wrap the label in parentheses

% \def\BibTeX{{\rm B\kern-.05em{\sc i\kern-.025em b}\kern-.08em
%     T\kern-.1667em\lower.7ex\hbox{E}\kern-.125emX}}

% move to main.tex
% \newcommand{\beginsupplement}{%
%         \setcounter{table}{0}
%         \renewcommand{\thetable}{A\arabic{table}}%
%         \setcounter{figure}{0}
%         \renewcommand{\thefigure}{A\arabic{figure}}%
%         \setcounter{section}{0}
%         \renewcommand{\thesection}{A\arabic{section}}%
% }

\beginsupplement
% \begin{document}
% \pagestyle{empty}

% \title{Conditional Residual Coding with Explicit-Implicit Temporal Buffering for Learned Video Compression\\ \textit{Supplementary Materials}}

% \author{%
% \IEEEauthorblockN{
% Yi-Hsin Chen\IEEEauthorrefmark{1} 
% \quad Kuan-Wei Ho\IEEEauthorrefmark{1} 
% \quad Martin Benjak\IEEEauthorrefmark{2}
% \quad Jörn Ostermann\IEEEauthorrefmark{2}
% \quad Wen-Hsiao Peng\IEEEauthorrefmark{1}
% %
% }
% %
% \\
% \IEEEauthorblockA{%
% \IEEEauthorrefmark{1}National Yang Ming Chiao Tung University, Taiwan \quad 
% \IEEEauthorrefmark{2}Leibniz Universität Hannover, Germany}
% }

% \maketitle
\maketitlesupplementary
% \thispagestyle{empty}
% \vspace{-3em}
\begin{figure*}[th]
    \centering
    \subfigure{
        \centering
        \includegraphics[width=.322\textwidth, trim= 0 0 55 50, clip]{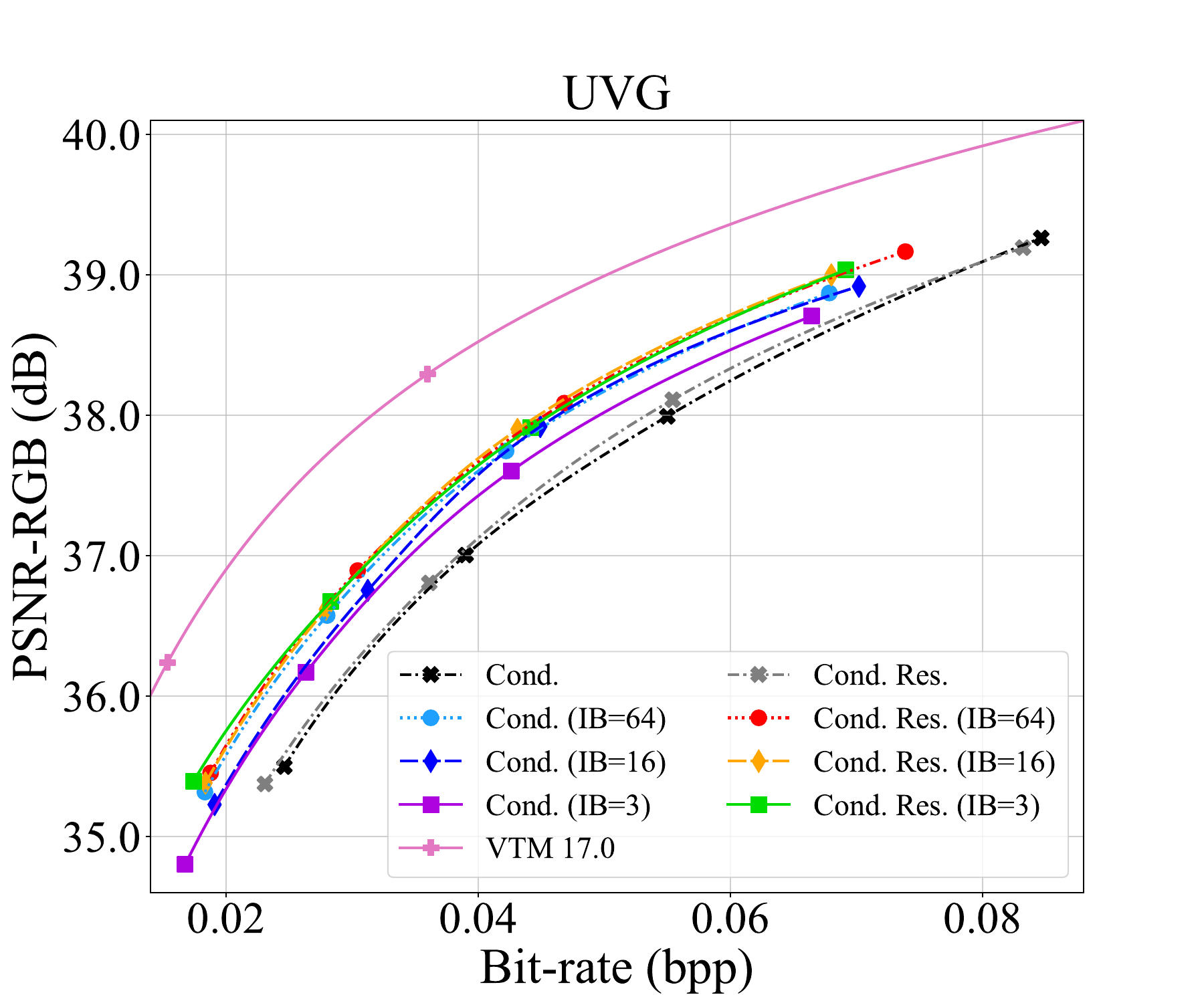}
        \label{fig:rd-a_supp}
        }
    \hspace{-3.6mm}
    \subfigure{
        \centering
        \includegraphics[width=.322\textwidth, trim= 0 0 55 50, clip]{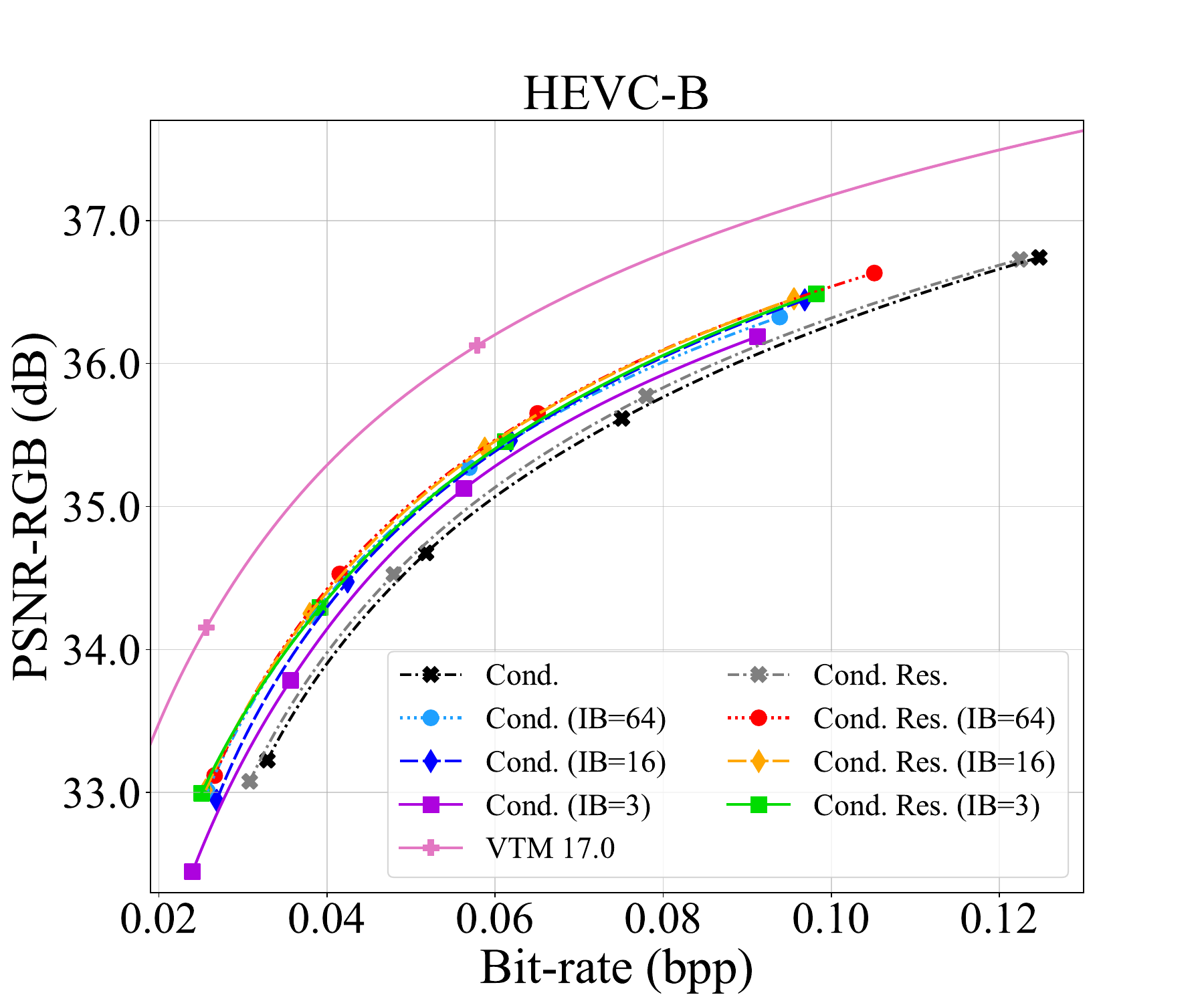}
        \label{fig:rd-b_supp}
    }
    \hspace{-3.6mm}
    % \vspace{-1.1mm}
    \vspace{-0.2mm}
    \subfigure{
        \centering
        \includegraphics[width=.322\textwidth, trim= 0 0 55 50, clip]{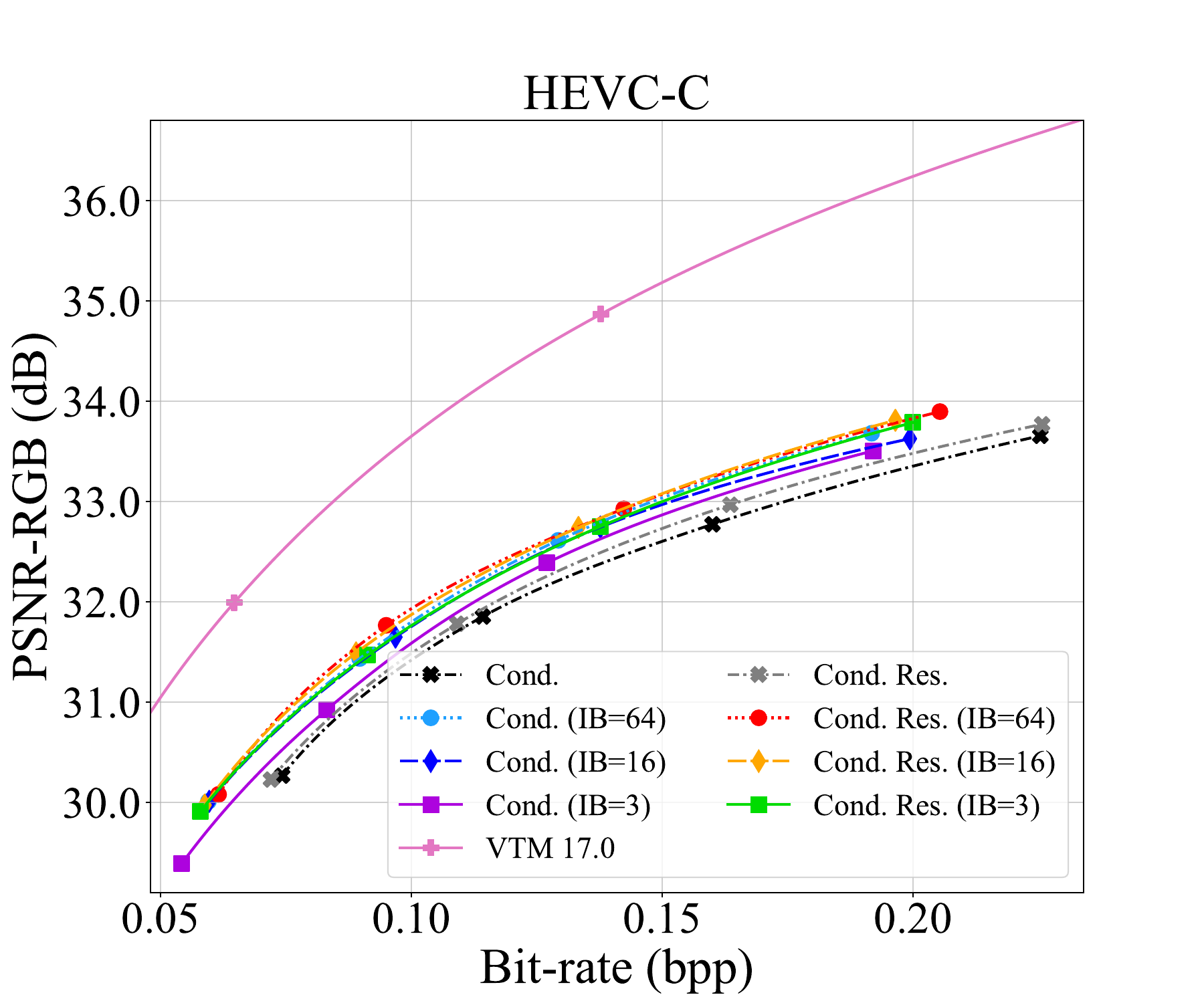}
        \label{fig:rd-c_supp}
    }
    \subfigure{
        \centering
        \includegraphics[width=.322\textwidth, trim= 0 0 55 50, clip]{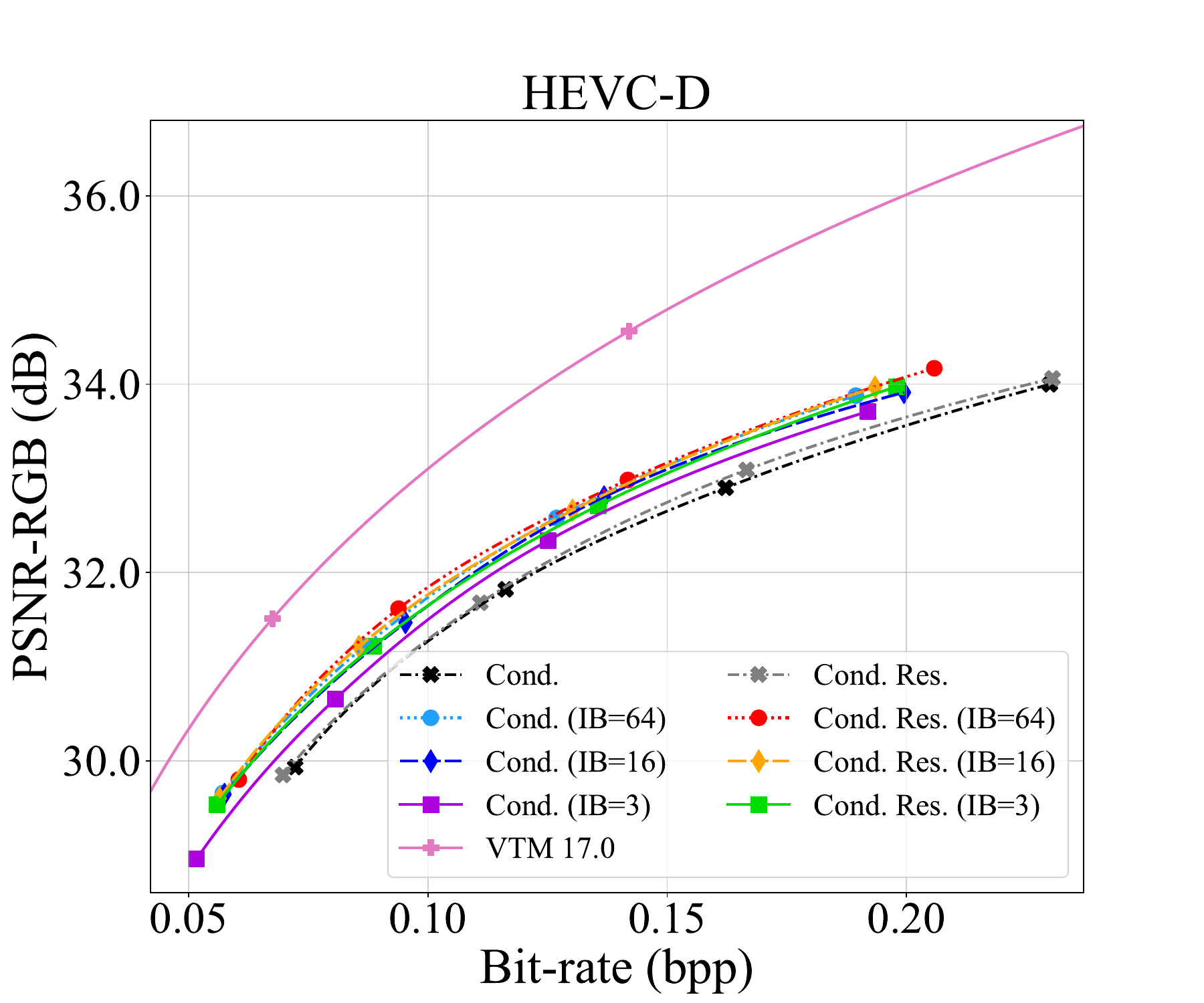}
        \label{fig:rd-d_supp}
        }
    \hspace{-2.8mm}
    \subfigure{
        \centering
        \includegraphics[width=.322\textwidth, trim= 0 0 55 50, clip]{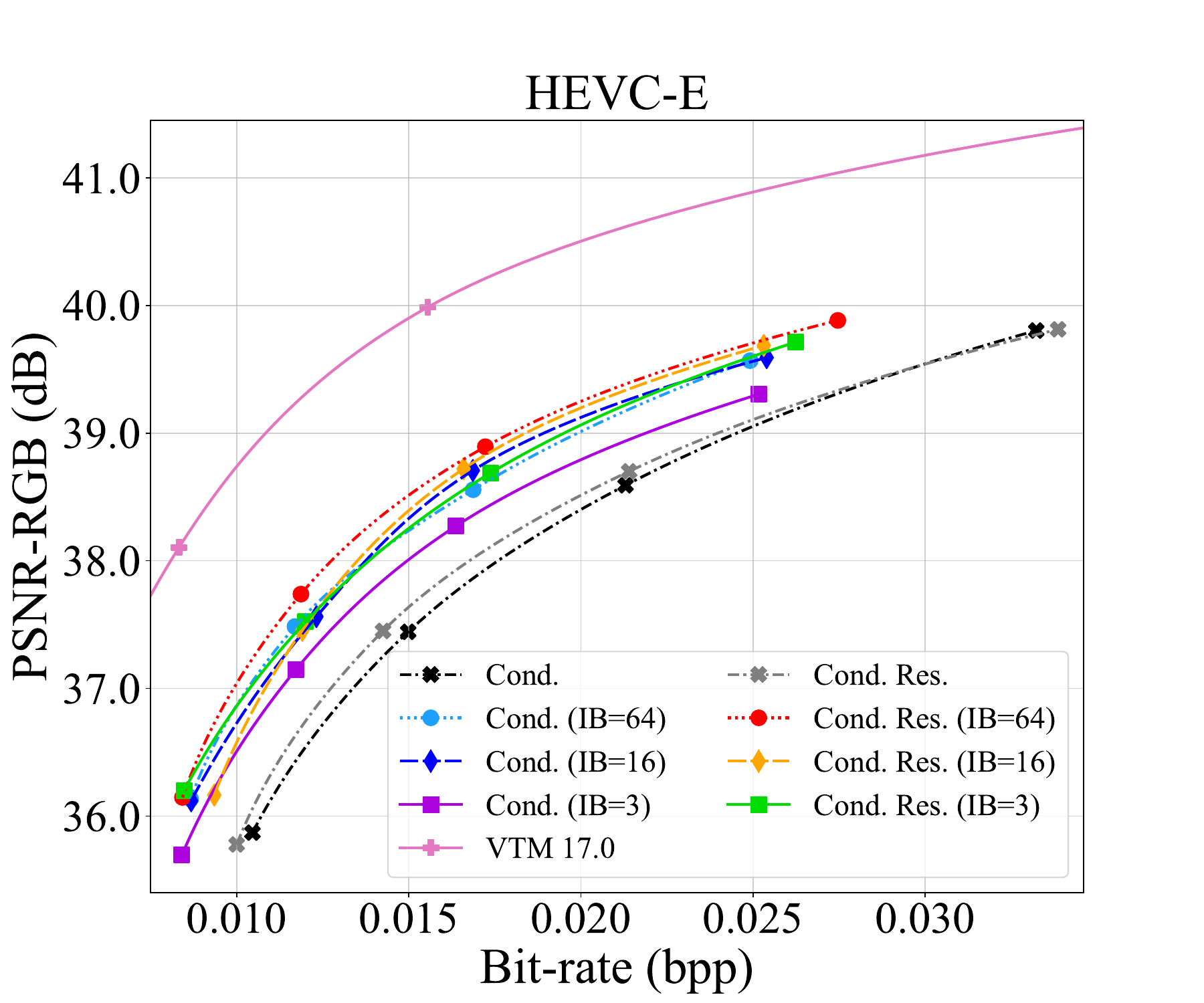}
        \label{fig:rd-e_supp}
    }
    \hspace{-3.6mm}
    \subfigure{
        \centering
        \includegraphics[width=.322\textwidth, trim= 0 0 55 50, clip]{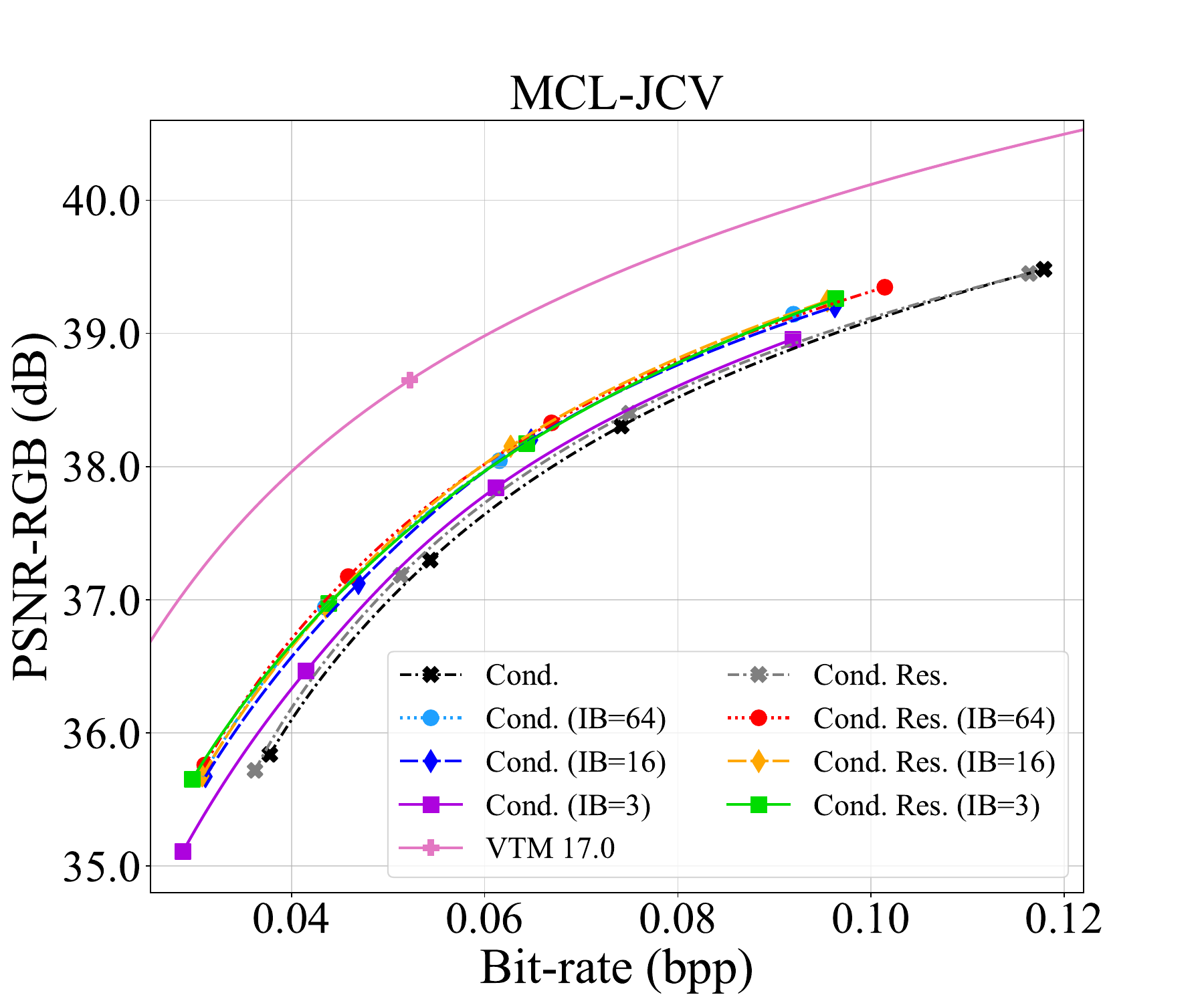}
        \label{fig:rd-f_supp}
    }
    % \vspace{-6.1mm}
    % \vspace{-4.5mm}
    \vspace{-4mm}
    \caption{Rate-distortion comparison with VTM~\cite{vtm}.}
    \vspace{-0.1cm}
    \label{fig:main_RD_supp}
\end{figure*}
% \vspace{-0.5em}
\begin{table*}[th]
\caption{BD-rate (\%) comparison with VTM~\cite{vtm} in terms of PSNR-RGB. The anchor is conditional coding without using implicit temporal information.} %The values in parentheses indicate the BD-rate changes relative to the codec type with $IB=64$.}
\label{table:main_RD_supp}
\centering
\setlength{\tabcolsep}{14pt}
\begin{tabular}{l|cccccc|c}
\toprule
                        & UVG & HEVC-B & HEVC-C & HEVC-D & HEVC-E & MCL-JCV & Average \\ 
% Only F_t
\midrule
    Cond.               & 0         & 0         & 0      & 0       & 0       & 0       & 0 \\
    Cond. (IB=64)       & -16.36    & -13.38    & -12.34 & -12.97  & -21.11  & -14.43  & -15.10  \\
    % Cond. (IB=16)       & -14.97 (+1.39) & -12.39 (+0.99) & -11.83 (+0.51) & -12.56 (+0.41) & -21.66 (-0.55) & -14.17 (+0.26)  & -14.60 (+0.50) \\
    Cond. (IB=16)       & -14.58    & -11.83    & -11.04 & -11.33  & -21.23  & -13.42  & -13.91  \\
    Cond. (IB=3)        & -9.26     & -7.96     & -6.44  & -6.93   & -14.56  & -7.11   & -8.71   \\
% both Ft xc
\midrule
    Cond. Res.          & -1.66     & -2.57     & -3.07  & -1.55  & -3.55    & -3.53  & -2.66  \\
    Cond. Res. (IB=64)  & -19.06    & -15.18    & -14.65 & -14.25  & -26.05  & -15.89  & -17.51  \\
    Cond. Res. (IB=16)  & -18.95    & -15.01    & -14.10 & -13.36  & -21.35  & -14.10  & -16.15  \\
    Cond. Res. (IB=3)   & -18.17    & -13.77    & -11.86 & -10.97  & -21.37  & -14.02  & -15.03  \\
\midrule
    VTM 17.0            & -39.62    & -39.09    & -47.72 & -36.62  & -54.91  & -43.71  & -43.61  \\
\bottomrule
\end{tabular}
\vspace{-3mm}
\end{table*}

This supplementary document presents a coding performance comparison between the latest standard video codec, VTM~\cite{vtm}, and the various codecs evaluated in the main paper with different buffer sizes.

% 
% \begin{itemize}
% \item Comparison of rate-distortion curves with traditional codec in Section~\ref{sec:traditional};

% Note that in this work, we use a simple learned video codec as a common base to fairly evaluate our hybrid buffering method. Therefore, comparisons with complex, state-of-the-art learned codecs or traditional codecs is not the focus of this study. Our method is simple yet effective and can likely be extended to advanced learned video codecs. As noted in the conclusion, future work includes extending our hybrid buffering scheme to state-of the-art learned video compression models.

% \end{itemize}

Following the recommendation from \cite{dcvc_dc}, we encode videos in YUV444 format. We use the \textit{encoder\_lowdelay\_vtm.cfg} of VTM~\cite{vtm} with the following parameters:

\vspace{0.1cm}

--c \{config file name\}

--InputFile=\{input file name\}

--InputBitDepth=8

--InputChromaFormat=444

--ChromaFormatIDC=444

--InternalBitDepth=10

--OutputBitDepth=8

--DecodingRefreshType=2

--FrameRate=\{frame rate\}

--FrameSkip=0

--SourceWidth=\{width\}

--SourceHeight=\{height\}

--FramesToBeEncoded=96

--Level=4.1

--IntraPeriod=32

--QP=\{qp\}

--BitstreamFile=\{bitstream file name\}

--ReconFile=\{reconstruction file name\}

% \FloatBarrier
\vspace{0.2cm}

Fig.~\ref{fig:main_RD_supp} and Table~\ref{table:main_RD_supp} present the rate-distortion and BD-rate comparisons, respectively. Note that this study employs a simple learned video codec as a common baseline to fairly evaluate our hybrid buffering method. Therefore, comparisons with complex state-of-the-art learned codecs or traditional codecs are not the focus. Our method is simple yet effective and has the potential to be integrated into advanced learned video codecs. As noted in the conclusion, future work includes extending our hybrid buffering approach to state-of-the-art learned video compression models.

% \bibliographystyle{IEEEbib}
% \bibliography{icme2025references}

% \end{document}

\end{document}